\documentclass[12pt]{article}

\font\sc=cmcsc10 scaled 1440

\def\rdots{\mathinner{\mkern1mu\raise1pt\vbox{\kern1pt\hbox{.}}\mkern2mu
   \raise4pt\hbox{.}\mkern2mu\raise7pt\hbox{.}\mkern1mu}}
\newcommand{\Z}{{\rm Z\kern-.35em Z}}
\newcommand{\bP}{{\rm I\kern-.15em P}}
\newcommand{\Q}{\kern.3em\rule{.07em}{.65em}\kern-.3em{\rm Q}}
\newcommand{\R}{{\rm I\kern-.15em R}}
\newcommand{\h}{{\rm I\kern-.15em H}}
\newcommand{\C}{\kern.3em\rule{.07em}{.65em}\kern-.3em{\rm C}}
\newcommand{\T}{{\rm T\kern-.35em T}}

\newcommand{\be}{\begin{equation}}
\newcommand{\ee}{\end{equation}}

\newcommand{\ve}{\varepsilon}

\newcommand{\pa}{\partial}

\newcommand{\La}{\Lambda}

\begin{document}

\openup 1.5\jot
\centerline{For the Quantum Heisenberg Ferromagnet, Some Conjectured
Approximations}

\vspace{1in}
\centerline{Paul Federbush}
\centerline{Department of Mathematics}
\centerline{University of Michigan}
\centerline{Ann Arbor, MI 48109-1109}
\centerline{(pfed@math.lsa.umich.edu)}

\vspace{1in}

\centerline{\underline{Abstract}}

We present some conjectured approximations for spin expectations in a
Quantum Heisenberg system.  The conjectures are based on numerical
experimentation, some theoretical insights and underpinning, and aesthetic
value.  We hope theoretical developments will follow from these ideas,
even leading to a proof of the phase transition (in three dimensions).

\vfill\eject

We organize this paper into three sections.  The first presents
preliminary definitions and the conjectures.  The second contains a
rigorous theoretical development of a useful framework for the system.
The final section introduces an ``average-field" structure that may lead
to an understanding, and hopefully a proof, of the conjectures.  

\bigskip

\centerline{{\sc I)\ \ \ \ The Conjectures.}}

We consider a lattice, $\La$, and the associated Quantum Heisenberg
Hamiltonian
\be	H = - \sum_{i \sim j} (I_{ij} - 1) 	\ee
where $I_{ij}$ interchanges the spins of the two neighboring sites $i$ and
$j$ in the lattice $\La$.  We let $p_i$ be the projection onto spin up at
site $i$.
\be	p_i = \frac 1 2 \ \big(\sigma_{zi} + 1\big).	\ee
We consider a state $\psi_0$ with spin up at sites in ${\cal S}_0$, and
spin down at the complementary sites.
\be
p_i \; \psi_0 = \left\{ 
\begin{array}{ll}
\psi_0 \ , & i \in {\cal S}_0 \\ 
 \\
0 \ , & i \not\in {\cal S}_0 
\end{array}
\right.
\ee
We assume there are $N$ spin ups,
\be	\# \ \{ {\cal S}_0 \} = N .    \ee

We let $\phi_\mu(i)$ be a solution of the lattice heat equation
\be	\frac \pa {\pa \mu} \ \phi_\mu(i) = ( \Delta \phi_\mu)(i)   \ee
with initial conditions
\be
 \phi_0(i) = \left\{
\begin{array}{ll}
1 \ , & i \in {\cal S}_0 \\ 
 \\
0 \ , & i \not\in {\cal S}_0 
\end{array}
\right.
\ee
We define
\be
\rho_\mu (i) = \frac {\phi^2_\mu (i)} {\phi^2_\mu(i) + (1-\phi_\mu(i))^2}
\ee
and
\be
<p_i>_\mu = \frac{\left< e^{-\mu H} \psi_0, \ p_i \; e^{-\mu H} \psi_0
\right>}
 		       {\left< e^{-\mu H} \psi_0, \  e^{-\mu H} \psi_0
\right>} \ \ .
\ee

\underline{Conjecture 1}.

\be   | <p_i >_\mu - \rho_\mu(i) | < c_d < 1 .	\ee

\underline{Conjecture 2}.

\be   | <p_i >_\mu - \phi_\mu(i) | < c_d < 1 .	\ee
In one-dimension the corresponding $c_1$ may be picked to be $\frac 1 2$,
according to our numerical studies.

\underline{Conjecture 3}.

\be   \lim_{\mu \rightarrow 0} \frac 1 \mu | <p_i >_\mu - \rho_\mu(i) | =
0 	\ee
the limit taken in $\ell^\infty(\Lambda)$, and convergence independent of
${\cal S}_0$ and $N$.  In one-dimension the $\mu$ in (11) may be replaced
by $\mu^{2-\ve}$.

Contrary to our earlier expectations (as presented in a previous version
of this note) the behavior of $< p_i >_\mu$ as $\mu$ becomes large is not
simple.  A better approximation than $\rho_\mu(i)$ or $\phi_\mu(i)$ when
$\mu$ is large is realized in $\phi_{\frac \mu 2} (i)$.  We present in our
next conjecture the result of our numeric study:

\underline{Conjecture 4}.

For $\mu \ge 4$ one has

\be   | <p_i >_\mu - \phi_{\frac \mu 2} (i) | < .1  \ . \ee
We have been specific with numbers in (12) to give the flavor of the
estimate's quality.  We have some theoretical understanding of the reason
$\phi_{\frac \mu 2} (i)$ is a good approximation to $< p_i >_\mu$, but we
do not discuss it in this note, restricting our attention to $\rho_\mu
(i)$ and $\phi_\mu (i)$ in later sections.

Implicit in all these estimates is a locality property of $< p_i>_\mu$.
We state a very weak form of this in the following conjecture.

\underline{Conjecture 5}.

For any $\ve > 0$, there is an $L_{\ve,\mu}$, such that $< p_i>_\mu$ is
determined within $\ve$ by knowledge of the spin configuration (as
specified in (3)) in a region within distance $L_{\ve,\mu}$ of site $i$,
i.e. changing spins outside this distance cannot effect $< p_i>_\mu$ by
more than $\ve$

Conjecture 5 is the most basic of our assertions, and should fit into some
very general theoretical framework.

\noindent
{\it Cave Adfirmationes}:  Most of the numerical investigation was on a
one-dimensional lattice.  (But also in periodic two-dimensional sets and
three-dimensional sets.)

\bigskip
\bigskip

\centerline{{\sc II)\ \ \ \ Some Simple Theory.}}

The Hilbert space of the system is naturally viewed as a direct sum
\[
{\cal H} = {\cal H}_0 \oplus {\cal H}_1 \oplus {\cal H}_2 \oplus \cdots
\oplus {\cal H}_{|\La|}
\]
where in ${\cal H}_N$ there are $N$ spin ups.  We write $H_N$ for $H$
restricted to ${\cal H}_N$.  The space ${\cal H}_N$ is an invariant
subspace of $H$, the set of $N$ spin waves.

We let $Q$ be an operator interchanging spin up and spin down, $Q$ a
unitary operator commuting with $H$.  $Q$ interchanges $H_N$ and
$H_{|\La|-N}$ as follows
\be
Q \left( \bigotimes_{i\in {\cal S}} \left( \begin{array}{c} 1 \\ 0
\end{array} \right)_i \ 
            \bigotimes_{j\not\in {\cal S}} \left( \begin{array}{c} 0 \\ 1
\end{array} \right)_j \right)
=  \bigotimes_{i\in {\cal S}} \left( \begin{array}{c} 0 \\ 1 \end{array}
\right)_i \  
            \bigotimes_{j\not\in {\cal S}} \left( \begin{array}{c}1 \\ 0
\end{array} \right)_j .
\ee

Vectors in ${\cal H}_N$ are described by symmetric functions on $N$
distinct lattice sites.  $f = f(.,. \ \ .,.)$ is associated to vectors in
${\cal H}_N$  as follows
\be
 f \longleftrightarrow \sum_{i_1,...,i_N} f(i_1, ..., i_N)
\bigotimes_{i\in\{i_i,...,i_N\}} 
\left( \begin{array}{c} 1 \\ 0 \end{array} \right)_i 
\bigotimes_{j \not\in\{i_i,...,i_N\}} 
\left( \begin{array}{c} 0 \\ 1 \end{array} \right)_j .
\ee
The sum in (14) is over distinct indices.

For $N > M$ there is a linear map from ${\cal H}_N$ to ${\cal H}_M$ called
$P_{N,M}$.  Let $f$ be in ${\cal H}_N$, then we define $P_{N,M} f$ in
${\cal H}_M$ by
\be
(P_{N,M} f)(i_1,...,i_M) = \sum_{i_{M+1},...,i_N} f(i_1,...,i_M \; , \;
i_{M+1},...,i_N)
\ee
$P_{N,M}$ commutes with $H$ and interlaces $H_N$ and $H_M$
\be	P_{N,M} H_N = H_M P_{N,M}  .  \ee 
If $2N \le |\La|$ then it is easy to show $P_{N,M}$ is onto.  The
preceding structure is related to the invariance of the system under
global rotations.

For $f$ in any ${\cal H}_N$ we define
\be	f_\mu \equiv e^{-\mu H} \ f \ .	\ee
Of course
\be	e^{-\mu H_M} P_{N,M}  = P_{N,M} e^{-\mu H_N}	.	\ee
In ${\cal H}_1$ , $f_\mu$ satisfies the heat equation
\be   \frac \pa {\pa \mu} f_\mu = -H_1 f_\mu = \Delta f_\mu \ .    \ee
So for $f$ in ${\cal H}_N$ we note the amusing fact that $P_{N,1} f_\mu$
satisfies the heat equation.

\bigskip

\centerline{{\sc III)\ \ \ \ An Average-Field Approximation.}}

Let $\psi_0$ be a state in ${\cal H}_N$, sharp in the spins, with spin up
at $i$ if $i \in {\cal S}_0$, spin down if $i\not\in {\cal S}_0$, ${\cal
S}_0$ a set of $N$ sites.
\be
\psi_0 \longleftrightarrow \bigotimes_{i\in {\cal S}_0} \left(
\begin{array}{c} 1 \\ 0 \end{array} \right)_i \  
            \bigotimes_{j\not\in {\cal S}_0} \left( \begin{array}{c}0 \\ 1
\end{array} \right)_j
\ee
and
\be	\psi_\mu = e^{-\mu H_N} \psi_0 \ .	\ee
We define
\be 	\phi_\mu = N \ P_{N,1} \ \psi_\mu	\ee
$\phi_\mu$ in $H_1$, satisfies the heat equation and
\be
 \phi_0(i) = \left\{ 
\begin{array}{ll}
1 \ , & i \in {\cal S}_0 \\ 
 \\
0 \ , & i \not\in {\cal S}_0 
\end{array}
\right.
\ee
We introduce an ``average-field"-like ``approximation" to $\psi_\mu$.
\be
\psi^{AP}_\mu \equiv \bigotimes_i \left( 
\begin{array}{c}
\phi_\mu(i) \\
  \\
1-\phi_\mu(i)   \end{array} \right)_i \ .
\ee

This approximation has two nice features.
\begin{description}
\item[1)] \ \ The approximation is ``invariant and $Q$".  That is, it is
$Q$ of the approximation obtained starting with $Q\psi_0$ instead of
$\psi_0$.
\item[2)]  \ \ The approximation is not sharp in spin wave number.  (It
does not lie in a single ${\cal H}_N$.)  But in a reasonable sense it
projects using $\{ P_{N,1} \}$ onto $\phi_{\mu}(i)$ in ${\cal H}_1$, that
does satisfy the heat equation.
\end{description}

We note that $\rho_\mu(i)$, from equation (7), is given by
\be
\rho_\mu(i) = \frac {\left< \psi^{AP}_\mu, p_i \psi^{AP}_\mu \right>}
			{\left< \psi^{AP}_\mu,  \psi^{AP}_\mu \right>} .
\ee
Thus our approximate wave function, the ``average-field" function (24)
yields the spin up probabilities of approximation 1, equation (9$'$).  The
``average-field" wavefunction satisfies the equation
\be
\frac	d {d\mu} \ \psi^{AP}_\mu = -H\; \psi^{AP}_\mu
\ee
in the limit of nearly constant $\phi_\mu(i)$.  We expect there is some
truth to the average field wavefunction; perhaps enough, so that its study
(objective genitive) leads towards a proof of the phase transition for
magnetization.    We note that $\phi_\mu(i)$ and $\rho_\mu(i)$ differ by
less than .16.  Both $\phi_\mu(i)$ and $\rho_\mu(i)$ are ``invariant under
$Q$" as approximations.

\end{document}